\documentclass[a4paper,11pt]{article}
\usepackage{lmodern}
\usepackage{pos}  

\newcommand{\e}{{\rm{e}}}
\newcommand{\img}{{\rm{i}}}

\title{$\Delta I =1/2$ process of $K\to\pi\pi$ decay on multiple ensembles with periodic boundary conditions}
\author{Masaaki Tomii (RBC \& UKQCD Collaborations)}
\emailAdd{masaaki.tomii@uconn.edu}
\affiliation{RIKEN BNL Research Center, Brookhaven National Laboratory, Upton, NY 11973, USA}
\affiliation{Physics Department, University of Connecticut, Storrs, CT 06269, USA}
\FullConference{The 41st International Symposium on Lattice Field Theory (LATTICE2024)\\
28 July - 3 August 2024\\
Liverpool, UK\\}

\abstract{
We present our preliminary results for the $\Delta I = 1/2$ matrix elements of $K\to\pi\pi$ decay and $\varepsilon'$, the measure of direct $CP$ violation in $K\to\pi\pi$, computed on multiple ensembles with periodic boundary conditions (PBC) at the inverse lattice spacings of $a^{-1} \approx 1.0$~GeV and 1.4~GeV.
The finer lattice ensemble is newly introduced as an extension to the first PBC calculation~\cite{RBC:2023ynh}, while the calculation on the coarser ensemble is updated with the approximately doubled statistics.
Our first attempt to take the continuum limit is also discussed with acknowledging potential significance of the $O(a^2)$ scaling violation on these coarse lattices.
}

\begin{document}
\maketitle

\section{Introduction}

Precise calculation of $K\to\pi\pi$ decay amplitudes has been a long-standing challenge for lattice QCD.
The measure of direct $CP$ violation, $\varepsilon'$, obtained from this calculation is expected to play a key role in testing the Standard Model (SM) and constraining the parameters of the Cabibbo-Kobayashi-Maskawa (CKM) matrix elements.
It is also expected to provide useful information for understanding the matter-antimatter imbalance in the present universe, while $CP$ violation within the SM appears insufficient to explain it.

The decays of kaons into two pions have two isospin processes, $\Delta I = 1/2$ and $\Delta I = 3/2$, and determining both isospin decay amplitudes enables us to give a prediction of $\varepsilon'$.
The $\Delta I=3/2$ amplitude was computed with sufficient precision about a decade ago~\cite{Blum:2011ng,Blum:2012uk,Blum:2015ywa}. 
The calculation of the $\Delta I = 1/2$ amplitude, on the other hand, needs to be improved to enable the SM prediction of $\varepsilon'$ to reach the experimental precision.
Since our first {\it ab-initio} lattice calculation of the $\Delta I = 1/2$ process~\cite{RBC:2015gro}, we have improved the calculation by introducing multiple two-pion operators and performing a step scaling as well as almost four times increasing the statistics~\cite{RBC:2020kdj}.  The introduction of an iso-singlet scalar operator as an additional two-pion operator for the $I=0$ channel played a crucial role in isolating the contamination from excited states.

The result for $\varepsilon'$ given in Ref.~\cite{RBC:2020kdj} is our best result as of now and reads Re$(\varepsilon'/\varepsilon) = 21.7(2.6)(6.2)$ $(5.0)\times10^{-4}$, where the errors, from left to right, correspond to statistical error, systematic error in the isospin limit and an estimate of electromagnetic and isospin-violating effects.  Although the result is in agreement with the experiment, Re$(\varepsilon'/\varepsilon) = 16.6(2.3)\times10^{-4}$, we need to improve the theory prediction to reach the experimental precision to avoid missing a potential opportunity to discover physics beyond the SM.

Besides the electromagnetic and isospin-violating effects~\cite{Cai:2018why,Christ:2021guf}, there are a few sources of significant systematic uncertainty that need to be addressed.
A particular error source is the finite lattice spacing effect as the simulation was carried out at a single and somewhat coarse lattice spacing, $a^{-1} \approx 1.4$~GeV.
While we employed G-parity boundary conditions (GPBC) in space in our first calculation of the $\Delta I =1/2$ process to realize the physical kinematics of $K\to\pi\pi$ with the ground two-pion state~\cite{RBC:2015gro,RBC:2020kdj}, we have demonstrated the feasibility of the same calculation with periodic boundary conditions (PBC), using an even coarser lattice, $a^{-1}\approx1.0$~GeV~\cite{RBC:2023ynh}.
This enables us to accelerate our calculation of $\varepsilon'$ using finer ensembles of domain wall fermions that were already generated with inverse lattice spacings up to $2.7$~GeV.

In this work, we calculate the $\Delta I =1/2$ $K\to\pi\pi$ amplitude using the PBC ensemble with the inverse lattice spacing $a^{-1}\approx1.4$~GeV and almost double the statistics on the coarser ensemble of $a^{-1}\approx1.0$~GeV as the first step toward the continuum limit.
These ensembles have the same fermion and gauge actions as the ones used in the GPBC calculation~\cite{RBC:2015gro,RBC:2020kdj,CK2024} except for the boundary conditions.  
The details of the ensemble parameters are found in Ref.~\cite{RBC:2020kdj} for the 1.4~GeV ensemble and Ref.~\cite{RBC:2023ynh} for the 1.0~GeV ensemble.


\section{Overview of computational details}

We use $2+1$-flavor M\"obius domain wall fermions (DWF) and the Iwasaki$+$DSDR gauge action.
The measurements are performed on lattices with $24^3\times64$ sites and $a^{-1}\approx1.0$~GeV and $32^3\times64$ sites with $a^{-1}\approx 1.4$~GeV.

The methodology of the measurement and analysis is largely unchanged from the previous work~\cite{RBC:2023ynh}.
We employ the all-to-all (A2A) quark propagator method~\cite{Foley:2005ac} with 2,000 low modes for the light quark and purely stochastic noise for the strange quark.
We also employ the all-mode averaging (AMA) technique~\cite{Blum:2012uh,Shintani:2014vja} to reduce the computational cost.
A prescription to implement AMA with the A2A method is described in Ref.~\cite{RBC:2023xqv}.
On the coarser lattice ($24^3$), the low modes are computed using the zM\"obius DWF action~\cite{Mcglynn:2015uwh} with a smaller 5D extension, $L_s=12$, and the sloppy quark propagators for both the light and strange quarks are computed with 400 CG iterations resulting in the residual $\sim5\times10^{-6}$.
The bias of such calculation is compensated by taking the difference from the exact calculation with roughly 10 times fewer configurations.  We employ the MADWF algorithm~\cite{Yin:2011np} to calculate the exact quark propagators with the M\"obius DWF action with $L_s=24$.
On the finer lattice ($32^3$), the M\"obius DWF action with $L_s=12$ is used for both sloppy and exact calculations.  Most of other details are the same as the $24^4$ lattice, but we stop the sloppy CG at 330 iterations on $32^3$.  The resulting residual is slightly larger than that on $24^3$, though the AMA correction is still sufficiently precise since we use, unlike the $24^3$ case, the same M\"obius DWF action for both sloppy and exact calculations.

We perform the hydrogen-like wave function smearing with a radius $\approx0.3$~fm for the pion and sigma interpolation operators and $\approx0.4$~fm for the kaon interpolating operator.
The two-pion operators are constructed by a product of pion interpolating operators with back-to-back spatial momentum and with a certain time separation of $\approx 0.6$~fm.
We introduce 4 different back-to-back momenta, $(0,0,0)$, $\frac{2\pi}{L}(0,0,1)$, $\frac{2\pi}{L}(0,1,1)$ and $\frac{2\pi}{L}(1,1,1)$ and project the operators to the $S$ wave and $I=0$.
In addition, we also include the sigma operator to well isolate the $I=0$ finite-volume two-pion states around the $f_0(500)$ resonance~\cite{RBC:2023ynh,RBC:2020kdj,RBC:2021acc}.

We compute the two-point functions of these two-pion operators
\begin{equation}
C_{ab}^{\rm2pt}(t,\delta_t)
= \langle O_a(t)O_b(0)^\dag\rangle - \langle O_a(t+\delta_t)O_b(0)^\dag\rangle,
\label{eq:pipi2pt}
\end{equation}
where $O_a$ stands for the two-pion operator labeled by $a$ and the second term subtracts unphysical contributions of constant terms such as the vacuum expectation values $\langle O_a\rangle\langle O_b^\dag\rangle$ and the effect of a single pion wrapping around the time boundary.

To obtain the $K\to\pi\pi$ matrix elements with an excited two-pion final state, we optimize the two-pion operators using the variational method~\cite{Luscher:1990ck,Blossier:2009kd}.  We solve the Generalized Eigenvalue Problem (GEVP) with a matrix of $\pi\pi$ two-point functions given in Eq.~\eqref{eq:pipi2pt}
\begin{equation}
C^{\rm2pt}(t,\delta_t)V_n(t,t_0,\delta_t)
= \lambda_n(t,t_0,\delta_t)C^{\rm2pt}(t_0,\delta_t)V_n(t,t_0,\delta_t),
\end{equation}
where $\lambda_n$ and $V_n$ are the (generalized) eigenvalue and eigenvector, respectively.  Here a trivial matrix multiplication between $C^{\rm2pt}(t,\delta_t)$ and $V_n(t,t_0,\delta_t)$ with respect to the suppressed $\pi\pi$ operator indices is understood.
The basis of $\pi\pi$ operators is determined by a re-basing technique described in Ref.~\cite{RBC:2023xqv}.
This is a minor update in the methodology from the previous work~\cite{RBC:2023ynh} but found to offer $\sim20$\% improvement in the statistical precision.

After optimizing the two-pion operator, we determine the finite-volume matrix elements
\begin{equation}
M_{n,i}^{\rm eff}(t_1,t_2,t_0,\Delta,\delta_t)
= C_{n,i}^{\rm3pt}(t_1,t_2,t_0,\Delta,\delta_t)
R^K(t_1)R_n^{\pi\pi}(t_2,t_0,\Delta,\delta_t),
\label{eq:efmlm}
\end{equation}
where 
\begin{align}
R^K(t) &= \sqrt{\frac{\e^{-m_K^{\rm eff}(t)t}}{C^K(t)}},
\\ 
R_n^{\pi\pi}(t,t_0,\Delta,\delta_t)
&= \sqrt{\frac{\lambda_n(t_0+\Delta,t_0,\delta_t)^{t/\Delta}}{V_n(t_0+\Delta,t_0,\delta_t)^\dag C^{\rm2pt}(t,\delta_t)V_n(t_0+\Delta,t_0,\delta_t)}},
\end{align}
with the two-point functions $C^K(t)$ and the corresponding effective mass $m_K^{\rm eff}(t)$ obtained from $C^K(t)$ and $C^K(t+1)$.
The $K\to\pi\pi$ three-point functions are defined by
\begin{equation}
C_{n,i}^{\rm3pt}(t_1,t_2,t_0,\Delta,\delta_t)
= \left\langle O_n^{(t_0+\Delta,t_0,\delta_t)}(t_1+t_2)
Q_i(t_1)O_K(0)^\dag\right\rangle,
\end{equation}
with the $\Delta S=1$ four-quark operators $Q_i$ and the kaon interpolating operator $O_K$.  To subtract the power divergence due to the mixing with the quark bilinear operator $\bar s\gamma_5d$, we perform the subtraction, $Q_i\to Q_i-\alpha_i\bar s\gamma_5d$, with $\alpha_i$ determined by a condition $\langle Q_i(t_1)O_K(0)\rangle=0$.
The two-pion operator is optimized for the $n$-th state, $O_n^{(t_0+\Delta,t_0,\delta_t)} = \sum_aV_{n,a}(t_0+\Delta,t_0,\delta_t)O_a$.
At sufficiently large values of $t_1$ and $t_2$, the effective matrix element of the $n$-th state in Eq.~\eqref{eq:efmlm} should plateau as long as the two-pion operator is well optimized by the GEVP procedure.


\section{Preliminary results}

\begin{figure}[tbp]
\begin{center}
\includegraphics[width=100mm]{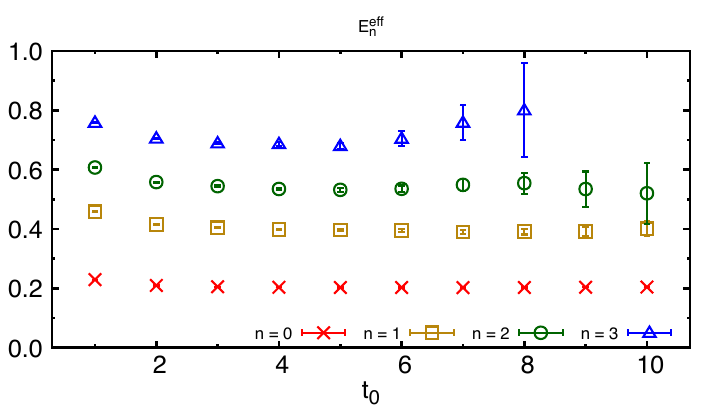}
\caption{
Two-pion effective energies $E_n^{\rm eff}(t_0+\Delta,t_0,\delta_t) = -\frac{1}{\Delta}\ln\lambda_n(t_0+\Delta,t_0,\delta_t)$ for the four low-lying states on the $32^3$ lattice plotted in lattice units.  The three low-lying states ($n=0,1,2$) are obtained from the GEVP with three re-based operators, while the result for $n=3$ is obtained from the GEVP with four re-based operators.  Here we choose $\Delta = 2$ and $\delta_t = 10$.
}
\label{fig:Eeff}
\end{center}
\end{figure}

Figure~\ref{fig:Eeff} shows an example of the effective energies for the four low-lying states on the $32^3$ ($a^{-1}\approx1.4$~GeV) lattice.  The first excited state ($n=1$) is extracted with 1\% precision and has the closest energy to the kaon mass, while the higher states up to $\sim1$~GeV are also resolved at $t_0\simeq 5$ where the plateau starts.

\begin{figure}[tbp]
\begin{center}
\includegraphics[width=100mm]{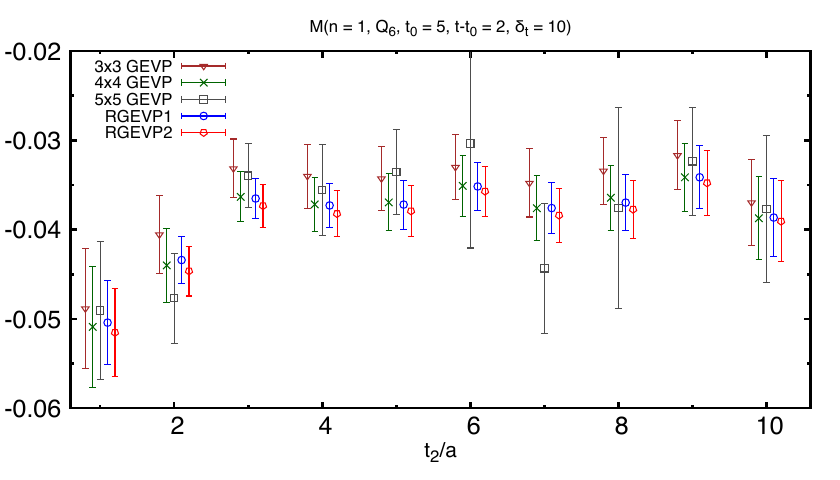}
\caption{
Unrenormalized effective matrix elements of $Q_6$ with the $n=1$ two-pion final state on the $32^3$ lattice plotted in lattice units.
}
\label{fig:efm}
\end{center}
\end{figure}

Figure~\ref{fig:efm} shows the effective matrix element of the unrenormalized four-quark operator $Q_6$ with the first-excited two-pion final state.  
We take a weighted average of $M_{1,6}^{\rm eff}(t_1,t_2,t_0,\Delta,\delta_t)$ in Eq.~\eqref{eq:efmlm} over $t_1$ using the covariance matrix in $t_1\ge4$, where no significant dependence on $t_1$ is observed.
The result of various GEVP bases are plotted. `$3\times3$ GEVP' corresponds to the GEVP {constructed with operators that create pions at rest, pions with spatial momentum $\pm\frac{2\pi}{L}(0,0,1)$ and the $\sigma$ at rest. The `$4\times4$ GEVP' includes an additional $\pi\pi$ operator with back-to-back momentum $\pm\frac{2\pi}{L}(0,1,1)$. The `$5\times5$ GEVP' further includes $\pi\pi$ operator with pions with momenta $\pm\frac{2\pi}{L}(1,1,1)$} as the fifth operator. `RGEVP1' and `RGEVP2' correspond to the $3\times3$ GEVP with three operators that are optimized for the three low-lying states beforehand by a re-basing prescription described in Ref.~\cite{RBC:2023xqv} using the original five operators.  While the re-basing depends on a few parameters, we show results of two different re-basing parameters as `RGEVP1' and `RGEVP2'.
The result of the `$5\times5$ GEVP' has larger errors than other GEVP results.  This typically happens when not all states considered in GEVP are well resolved~\cite{RBC:2023ynh,RBC:2023xqv}.
The results of `RGEVP1' and `RGEVP2' are roughly 20\% better resolved than the `$3\times3$ GEVP' and the `$4\times4$ GEVP'.
This indicates that pre-optimizing the two-pion operators with a few additional operators could offer better statistical precision in matrix elements.

These matrix elements need to be multiplied by the Lellouch-L\"uscher factor~\cite{Lellouch:2000pv} to exponentially suppress their finite volume effects.
On the $24^3$ and $32^3$ lattices in this study, only the two lowest energy states are in the elastic region where the Lellouch-L\"uscher factor is strictly valid.
By multiplying the lattice matrix elements by this factor, we obtain the $K\to\pi\pi$ matrix elements with exponentially suppressed finite volume effects for each two-pion finite-volume energy.
Without tuning the lattice volume, the finite-volume energies of two pions do not coincide with the kaon mass.  To realize the physical kinematics of $K\to\pi\pi$ we linearly interpolate the matrix elements to the on-shell point with respect to two-pion energy.
While this assumption may cause a systematic error, we estimated it to be much smaller than the statistical error on the $24^3$ lattice because the energy of the first-excited two-pion state ($n=1$) is only 6\% larger than the kaon mass~\cite{RBC:2023ynh}.
On the $32^3$ lattice, the $n=1$ two-pion energy is about 11\% larger than the kaon mass and this could enhance the systematic error, but we postpone the estimation of this effect until the full paper.

\begin{figure}[tbp]
\begin{center}
\begin{tabular}{c}
\includegraphics[width=70mm]{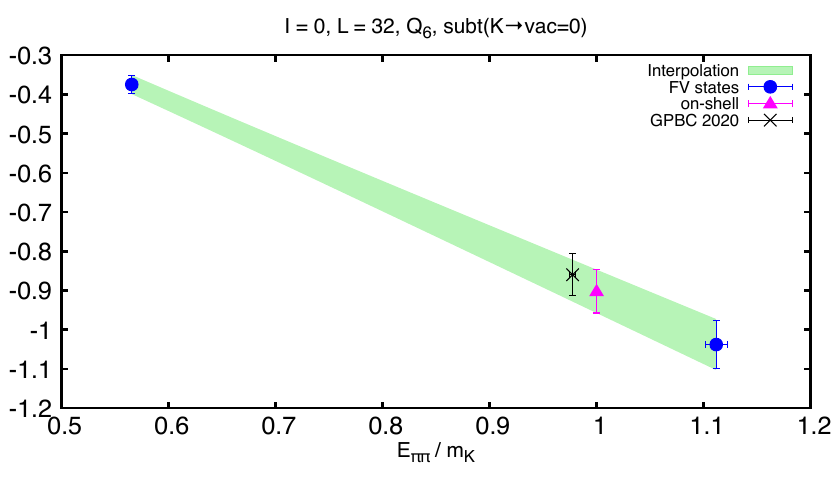}
\end{tabular}
\hfill
\begin{tabular}{c}
\includegraphics[width=70mm]{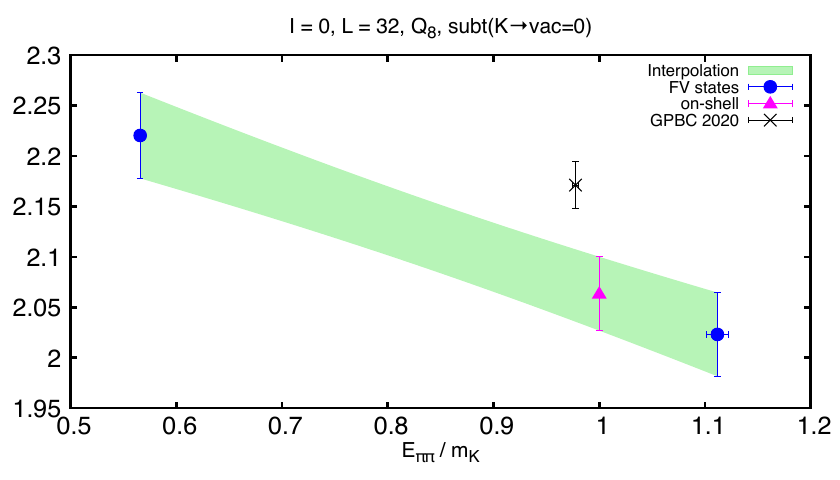}
\end{tabular}
\caption{
Interpolation of unrenormalized matrix elements of $Q_6$ (left) and $Q_8$ (right) multiplied by the Lellouch-L\"uscher factor to the on-shell point on the $32^3$ lattice plotted in lattice units.
The corresponding GPBC results in Ref.~\cite{RBC:2020kdj} are also plotted.
}
\label{fig:interp_MEs}
\end{center}
\end{figure}

Figure~\ref{fig:interp_MEs} shows the interpolation of the unrenormalized matrix elements of $Q_6$ and $Q_8$ with the Lelloush-L\"uscher factor multiplied to the on-shell point.
Although they are bare matrix elements, we can compare the results with those from GPBC calculation~\cite{RBC:2020kdj} because the $32^3$ lattice is generated with the same lattice action and with the same bare coupling as the GPBC calculation.  The only difference is boundary conditions.
Therefore we also plot the corresponding matrix elements from the GPBC multiplied by its individual Lellouch-L\"uscher factor.
The PBC and GPBC results should be consistent up to the remaining finite volume effect which is exponentially suppressed.
While most of the matrix elements are in good agreement with GPBC, some tension is observed for those of $Q_7$ and $Q_8$, which are electroweak penguin operators with the left-right chirality.

We renormalize these matrix elements with the same procedure as in Ref.~\cite{RBC:2023ynh,RBC:2020kdj}.  We employ the RI/SMOM schemes as an intermediate scheme, in which we nonperturbatively perform scale evolution to $\mu=4.0$~GeV using the step-scaling procedure~\cite{Arthur:2010ht}.
After the scale evolution, we perturbatively convert the renormalization scheme to $\rm\overline{MS}$.
The only difference in the detailed procedure from the previous PBC work~\cite{RBC:2023ynh} is that the present work performs renormalization after interpolating the matrix elements to the physical kinematics, although we renormalized the matrix elements before the interpolation in the previous work. 
It is easy to show that the on-shell renormalized matrix elements do not depend on the order of interpolation and renormalization as long as the interpolation is done with two data points.

\begin{figure}[tbp]
\begin{center}
\begin{tabular}{c}
\includegraphics[width=70mm]{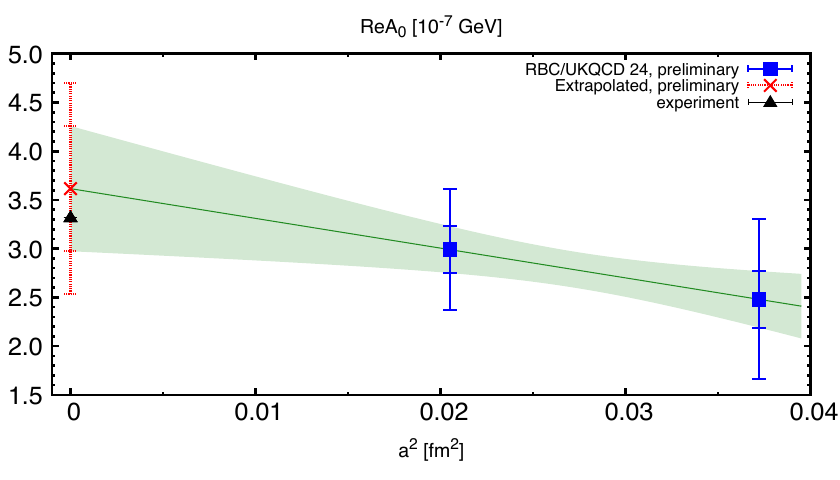}
\end{tabular}
\hfill
\begin{tabular}{c}
\includegraphics[width=70mm]{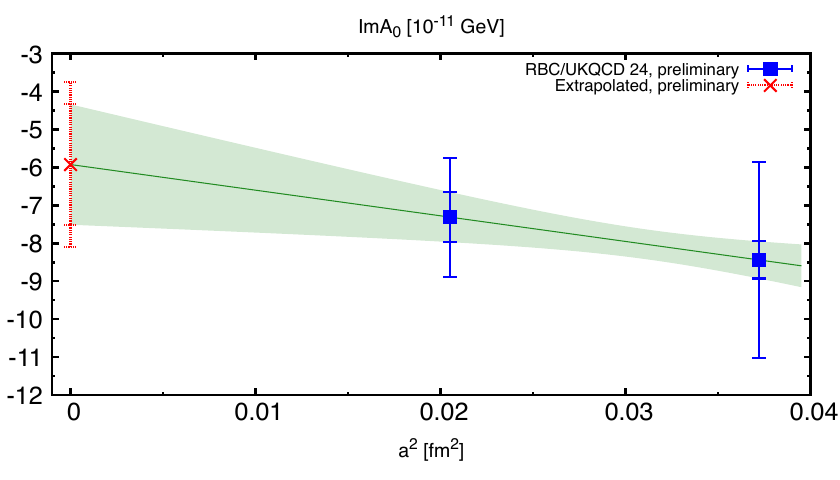}
\end{tabular}
\caption{
Continuum extrapolation of the real (left) and imaginary (right) parts of $A_0$.
The smaller and larger error bars of lattice results represent the statistical and total errors, respectively.
Since all systematic errors are mutual for the two lattice spacings with an exception of the finite lattice spacing error, which should be excluded from the error on the data points when doing the continuum extrapolation, the extrapolation is performed with statistical the error only and then the systematic errors are added in quadrature to both extrapolated and finite-lattice values.  The experimental value of Re$A_0$ is also plotted.
}
\label{fig:extrap_ampl}
\end{center}
\end{figure}

Combining the renormalized matrix elements $M_i^{\rm\overline{MS}}(\mu)$ with the Wilson coefficients $y_i^{\rm\overline{MS}}(\mu)$ and $z_i^{\rm\overline{MS}}(\mu)$ in three-flavor theory given in Ref.~\cite{Buras:1993dy,Buchalla:1995vs},
we calculate the $\Delta I=1/2$ $K\to\pi\pi$ amplitude $A_0$,
\begin{equation}
A_0 = \frac{G_F}{\sqrt2}V_{us}^*V_{ud}\sum_i
[z_i^{\rm\overline{MS}}(\mu) + \tau y_i^{\rm\overline{MS}}(\mu)]
M_i^{\rm\overline{MS}}(\mu),
\end{equation}
where we define the Fermi constant $G_F$, the CKM matrix elements $V_{q'q}$ connecting up-type ($q'$) and down-type ($q$) quarks and their ratio $\tau = -V_{ts}^*V_{td}/V_{us}^*V_{ud}$.
We perform the same estimation of the systematic errors as in Ref.~\cite{RBC:2023ynh}.
In Figure~\ref{fig:extrap_ampl} we show $A_0$ versus $a^2$ and its extrapolation to the continuum limit $a\to0$, assuming $O(a^2)$ scaling.
The smaller error bars represent the statistical error, while the larger ones correspond to the total error with systematic errors added in quadrature.
The total error on the continuum limit includes the systematic error due to the $O(a^2)$ scaling violation, which is estimated by taking the difference in the central value from the continuum extrapolation with an assumption of $O(a^4)$ scaling, instead of the finite lattice spacing error included in the finite-lattice values.
Besides the discretization or scaling-violation error, the error from perturbative truncation of Wilson coefficients is the most significant (12\%).

Finally we calculate the continuum limit of the measure of direct $CP$ violation relative to that of indirect $CP$ violation in the isospin limit,
\begin{equation}
\frac{\varepsilon'}{\varepsilon} = 
\frac{\img\omega\e^{\img(\delta_2-\delta_0)}}{\sqrt2\varepsilon}
\left[
\frac{{\rm Im}A_2}{{\rm Re}A_2}
- \frac{{\rm Im}A_0}{{\rm Re}A_0}
\right],
\end{equation}
where we define the $\Delta I=3/2$ amplitude $A_2$, the isospin-$I$ $\pi\pi$ phase shifts $\delta_I$, and $\omega = {\rm Re}A_2/{\rm Re}A_0$.
We use the values of Re$A_0$ and Re$A_2$ from the experiments and Im$A_2$ from RBC/UKQCD's earlier work~\cite{Blum:2015ywa} with a small change due to the PDG update on the PDG value of $\tau$, following the procedure in our previous works~\cite{RBC:2023ynh,RBC:2020kdj}.
Our preliminary result reads Re$(\varepsilon'/\varepsilon) = 17.5(6.8)(4.9)(5.0)$, which is in good agreement with the experiment Re$(\varepsilon'/\varepsilon)_{\rm exp} = 16.6(2.3)$ and the previous GPBC calculation, Re$(\varepsilon'/\varepsilon)_{2020} = 21.7(2.6)(6.2)$ $(5.0)\times10^{-4}$.  The three errors in lattice calculations, from left to right, correspond to the statistical and systematic errors in the isospin limit, and an estimate of the electromagnetic/isospin-violating corrections.

\begin{figure}[tbp]
\begin{center}
\begin{tabular}{c}
\includegraphics[width=90mm]{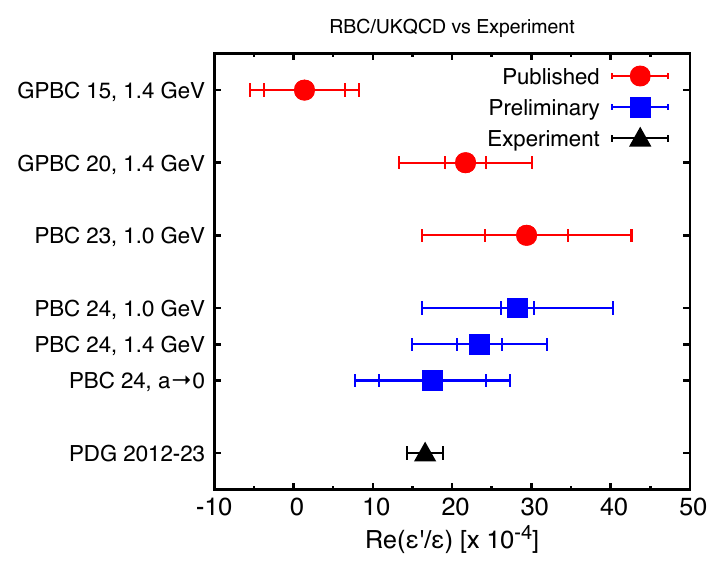}
\end{tabular}
\caption{
History of RBC/UKQCD calculation of Re$(\varepsilon'/\varepsilon)$ compared with the world average of experiments.
The smaller and larger error bars of lattice results represents the statistical and total errors, respectively.
}
\label{fig:history_epsp}
\end{center}
\end{figure}

\section{Summary and outlook}

In this work, we calculate the $\Delta I = 1/2$ $K\to\pi\pi$ amplitude and $\varepsilon'$, the measure of direct $CP$ violation on two lattice ensembles of $a^{-1}\approx1.0$~GeV and $a^{-1}\approx1.4$~GeV with periodic boundary conditions, as an extension of our earlier PBC calculation~\cite{RBC:2023ynh} where we calculated on the coarser lattice of $a^{-1}\approx 1.0$~GeV.
We present, for the first time, the preliminary result for the continuum extrapolation of the amplitude and $\varepsilon'$.
In Figure~\ref{fig:history_epsp} we summarize the results for Re($\varepsilon'/\varepsilon$) in our earlier calculations (circles) and in this work (squares, preliminary) along with the world average of experiments.
The result in the present work is consistent with the experimental results and our earlier lattice results, while the lattice calculation is still desired to be improved.

This work represents an important first step towards reducing the finite lattice spacing error, which is one of the most significant systematic errors estimated in our earlier works~\cite{RBC:2023ynh,RBC:2020kdj}, as it calculates the $\Delta I =1/2$ amplitudes with multiple lattice spacings for the first time.
While we present the first continuum extrapolations of the amplitude and $\varepsilon'$ with an estimation of the $O(a^2)$ scaling violation error (11\% for Im$A_0$), this error estimation {is somewhat uncertain since we use only two lattices with relatively large spacings}.
We plan to continue calculations on existing M\"obuis DWF lattice ensembles with PBC with larger inverse lattice spacings up to 2.7~GeV.
The next generation of $K\to\pi\pi$ calculation with these finer lattices will significantly reduce the discretization error.

Another significant source of systematic error within the isospin limit is the perturbative truncation of Wilson coefficients ($\sim12$\% for $A_0$).
This error cannot be reduced by step scaling because Wilson coefficients need to be matched between three- and four-flavor theories below the charm threshold, where NLO perturbation theory could be quite uncertain.
This error could be reduced in the near future either by an NNLO perturbative calculation or by nonperturbative matching between three- and four-flavor theories~\cite{Tomii:2020smd}.

Introduction of electromagnetic and isospin-violating corrections is also a very important piece for achieving the experimental precision of $\varepsilon'$.
Theoretical studies are underway~\cite{Cai:2018why,Christ:2021guf} and numerical calculation will be performed in the future.

\acknowledgments
This work was supported in part by the DOE Office of Science Early Career Award DE-SC0021147 and Laboratory Directed Research and Development (LDRD No. 23-051) of BNL and RIKEN-BNL Research Center.
The numerical calculations were carried out with the USQCD resources funded by the US DOE at BNL and JLab.
I thank the members of the RBC and UKQCD collaborations for fruitful discussions.
I thank T.~Blum for his careful read of the manuscript.

\end{document}